\documentclass [twocolumn,superscriptaddress,floatfix,showpacs,draft] {revtex4}

\begin{document}

\title{Comment on ``Sodium Pyroxene NaTiSi$_2$O$_6$: Possible Haldane Spin-1
Chain System''}

\author{S.V.~Streltsov}
\affiliation{Institute of Metal Physics, S.Kovalevskoy St. 18, 620219 Ekaterinburg GSP-170, Russia}
\affiliation{II. Physikalisches Institut, Universit$\ddot a$t zu K$\ddot o$ln,
Z$\ddot u$lpicher Stra$\ss$e 77, D-50937 K$\ddot o$ln, Germany}
\email{streltsov@optics.imp.uran.ru}

\author{O.A.~Popova}
\affiliation{Institute of Metal Physics, S.Kovalevskoy St. 18, 620219 Ekaterinburg GSP-170, Russia}
\affiliation{Ural State Technical University, Mira St. 19, 620002 Ekaterinburg, Russia}

\author{D.I.~Khomskii}
\affiliation{II. Physikalisches Institut, Universit$\ddot a$t zu K$\ddot o$ln,
Z$\ddot u$lpicher Stra$\ss$e 77, D-50937 K$\ddot o$ln, Germany}


\maketitle
A rather rare phenomenon of an opening of the spin gap in transition metal
oxides was observed in pyroxene NaTiSi$_2$O$_6$ and was interpreted
as a formation of singlet Ti-Ti dimers~\cite{Isobe-02,Hikihara-04}.
However, in the recent Letter ~\cite{Popovich-04} the authors
challenged this picture. On the basis of spin polarized
GGA calculations they argued that with decrease of the temperature this
compound evolves into a Haldane phase, characterized by formation
of one-dimensional S=1 chains. 

This novel interpretation, however, is highly questionable. The
authors claim that ``direct overlap between 3d orbitals centered on {\it closer Ti} 
ions, ..., indicates that two electrons of {\it the same spin}, occupying those states, 
are shared by two Ti ions'' (cursive ours). This picture however contradicts
typical situation for shorts metal-metal bonds in insulating solids,
especially those of spin $S=1/2$ ions.

Most probably the defect lies in the neglect of electronic
correlations in the calculation method used in~\cite{Popovich-04}. 
NaTiSi$_2$O$_6$ is known to be strong Mott insulator with the energy gap
close to 2~eV~\cite{Isobe-05}, whereas the calculations of
Ref.~\onlinecite{Popovich-04} lead to very small energy gaps 0.2-0.3~eV.

In order to take into account strong electronic correlations
in Ti-$3d$ shell, we performed for the LT phase (T=100~K) 
of NaTiSi$_2$O$_6$ the LDA+U calculations~\cite{Anisimov-91}, which was
proven very successful in similar 
cases~\cite{Leonov-05,Streltsov-05,Seidel-03}

The calculation
scheme was realized in the framework of the linear muffin-tin orbitals
method~\cite{Andersen-84}. Crystal structure parameters were 
taken from Ref.~\onlinecite{Redhammer-03}.
The values of on-cite Coulomb interaction $U=3.3$~eV and Hund's rule exchange
$J_H=0.8$~eV parameters for Ti-$3d$ shell were obtained in constrained supercell
calculation~\cite{Streltsov-05}.

In contrast to Ref.~\onlinecite{Popovich-04}, the fully antiferromagnetic state (AFM) was found
to have lowest total energy (see Tab.1). The total energy of  F+AF state 
(ferromagnetically coupled antiferromagnetic dimers on short Ti-Ti bonds) is
just a little bit larger. This indicates that the coupling between dimers is
pretty small, $J^{inter} = 7$~K. On the contrary, the exchange interaction
within Ti dimers is {\it antiferromagnetic} and rather strong: from the comparison
of total energies of totally AFM and AF+F
states we find $J^{intra} = 626$~K. Direct calculation of exchange integrals using 
the scheme of Ref.~\onlinecite{Katsnelson-00} gives very similar values for J's.
Thus, our results confirm the formation of singlet Ti-Ti dimers in the LT phase 
of NaTiSi$_2$O$_6$, as originally proposed ~\cite{Isobe-02,Hikihara-04}.
The spin-gap would be even bigger then the one found in the experiment,
$\Delta_{exp} \sim$500~K~\cite{Isobe-02}.  The obtained value of the 
band gap $E_g= 1.71$~eV is also
of the right order~\cite{Isobe-05}.
Thus our calculations give a quite reasonable description of
the main physical properties of NaTiSi$_2$O$_6$, very different from 
that proposed in Ref.~\onlinecite{Popovich-04}.

\begin{table*}
\centering \caption{Total energies (per dimer), values of magnetic moment and band 
gap for different magnetic solutions. Ti ions forming dimers with short 
Ti-Ti distances are connected by ``$-$''. }
\vspace{0.2cm} \label{SplitTable}
\begin{tabular}{lcccc}
\hline
\hline
 & Magnetic ordering & Total energy (meV) & Spin moment ($\mu_B$) &  Band gap (eV)\\
\hline
AFM  & $\uparrow-\downarrow$~~~$\uparrow-\downarrow$ & 0   & 0.91 & 1.71 \\
F+AF & $\uparrow-\downarrow$~~~$\downarrow-\uparrow$ & 0.6 & 0.91 & 1.66 \\
AF+F & $\uparrow-\uparrow$~~~$\downarrow-\downarrow$ & 54  & 0.92 & 1.5 \\
FM   & $\uparrow-\uparrow$~~~$\uparrow-\uparrow$     & 53  & 0.93 & 1.35 \\
\hline
\end{tabular}
\end{table*}

In conclusion, we stress that by neglecting correlation effects
~\cite{Popovich-04} one gets incorrect description of magnetic properties of
NaTiSi$_2$O$_6$, which definitely Mott insulator.
Our calculations demonstrate the formation of singlet
Ti-Ti dimers in LT phase of NaTiSi$_2$O$_6$ driven by corresponding orbital 
ordering~\cite{Hikihara-04}, which strongly disagrees
with the results and interpretations of Ref.~\onlinecite{Popovich-04}.

This work is supported by the projects DAAD A/05/01025/325,
INTAS 05-109-4727, RFFI-04-02-16096, NWO 047.016.005, 
Ural branch of RAS through the YS support program, and SFB 608.


\begin{thebibliography}{21}
\bibitem{Isobe-05} The color of NaTiSi$_2$O$_6$ sample is grayish green
(M. Isobe, private commun.). It means that the band gap is of order of 
2~eV.
\end{thebibliography}
\end{document}